\documentclass{article}
\usepackage{amssymb}

\begin{document}

\centerline{The average determinant of the reduced density matrices for
each qubit} \centerline{as a global entanglement measure }
\centerline{Dafa
Li}
\centerline{Department of mathematical sciences, Tsinghua University,
Beijing 100084 CHINA} \centerline{email: lidafa@tsinghua.edu.cn}

Abstract. In this paper, we propose the average determinant of reduced
density matrices for each qubit as a global entanglement measure. By means
of the properties of reduced density matrices, we can investigate the
present measure. We propose a decomposition law for the present measure,
demonstrate that the present measure just measures the average mixedness for
each qubit and the average 1-tangle, and indicate that for $n$-qubit W state
and Dicke states, the average mixedness for each qubit and 1-tangle almost
vanish for the large number of qubits. We also point out that for two
qubits, the present measure is just the square of the concurrence while for
three qubits, the present measure is the sum of the 3-tangle and the twice
the average 2-tangle.


\section{Introduction}

Quantum entanglement is a unique quantum mechanical resource \cite{Nielsen}.
Entanglement takes a critical role in quantum information processing and
quantum computing, for example in quantum teleportation, quantum superdense
coding, quantum error correction coding, quantum cryptography, quantum
metrology, and quantum key distribution.

Many efforts have been made to study measures of quantum entanglement \ and
classification of entanglement. In previous papers, the following measures
of entanglement are proposed: concurrence, 1-tangle, 2-tangle, 3-tangle,
Meyer-Wallach measure of global entanglement, entanglement of formation,
linear entropy, negativity, von Neumann entanglement entropy, and so on \cite%
{Nielsen,Coffman, Vidal, Ou}. The entanglement classification was explored
via LU, local operations and classical communication (LOCC), and Stochastic
LOCC (SLOCC) \cite{Linden, Grassl, Carteret, Sudbery, Dur,
Verstraete,Li-prl, Li-pra}. For example, three qubits are partitioned into
six SLOCC\ equivalence classes, two of which are GHZ and W classes which are
genuinely entangled \cite{Dur}.

Meyer and Wallach proposed the measure of global entanglement for pure
states of n qubits via the norm-squared of the wedge product of the two
vectors $|u^{(k)}\rangle $ and $|v^{(k)}\rangle $ \cite{Meyer}.
Meyer-Wallach measure is studied in \cite{Roya, Brennen, Scott, Pratik,
Haug, Victor}, applied to track the evolution of entanglement during a
quantum computation, and used for quantum phase transition \cite{Meyer, Roya}%
.

In this paper, we propose the average determinant of reduced density
matrices for each qubit as a global entanglement measure. Using the
properties of the reduced density matrices, we propose a decomposition law
for the present measure.

\section{The average determinant of the reduced density matrices for each
qubit as a global entanglement measure}

In this paper, for the normalized pure state $|\psi \rangle _{1\cdots n}$ of
$n$\ qubits, we propose
\begin{equation}
E_{AD}(|\psi \rangle )=\frac{4}{n}\sum_{i=1}^{n}\det \rho _{i}
\end{equation}%
as a global entanglement, where $\rho _{i}$ is the reduced density matrix
for the $ith$ qubit obtained by tracing over the rest qubits and $\det \rho
_{i}$ is the determinant of $\rho _{i}$. It is known that $0\leq \det \rho
_{i}\leq 1/4$ and $\det \rho _{i}$ is a polynomial of degree 4 and LU
invariant. $E_{AD}$ is just the average determinant of the reduced density
matrices for each qubit.

A calculation yields that for two qubits, $E_{AD}$\ is just the square of
the concurrence (ref. Appendix C) while for three qubits, $E_{AD}$ is the
sum of 3-tangle and the double of the average 2-tangles (ref. Appendix D).

We compute $E_{AD}$ for some states in Appendix A.

\subsection{Decomposition law}

Let $\rho $\ be the density matrix of one-qubit state $\alpha |0\rangle
+\beta |1\rangle $. Then, $\det \rho =0$. From this, we can define $E_{AD}=0$
for one-qubit state $\alpha |0\rangle +\beta |1\rangle $.

Proposition 1 (Decomposition law). Let $|\psi \rangle _{1\cdots n}=|\phi
\rangle _{i_{1}\cdots i_{k}}\otimes |\varphi \rangle _{j_{1}\cdots j_{\ell
}} $, where\ $k+\ell =n$ and$\ |\phi \rangle _{i_{1}\cdots i_{k}}$ and $%
|\varphi \rangle _{j_{1}\cdots j_{\ell }}$ are normalized. Then,
\begin{equation}
E_{AD}(|\psi \rangle _{1\cdots n})=\frac{kE_{AD}(|\phi \rangle _{i_{1}\cdots
i_{k}})+\ell E_{AD}(|\varphi \rangle _{j_{1}\cdots j_{\ell }})}{n}.
\label{linear-1}
\end{equation}

Proof. Let $\rho _{1\cdots n}=|\psi \rangle _{1\cdots n}\langle \psi |$, $%
\sigma _{i_{1}\cdots i_{k}}=|\phi \rangle _{i_{1}\cdots i_{k}}\langle \phi |$%
, and $\upsilon _{j_{1}\cdots j_{\ell }}=|\varphi \rangle _{j_{1}\cdots
j_{\ell }}\langle \varphi |$. Then,
\begin{equation}
\rho _{1\cdots n}=\sigma _{i_{1}\cdots i_{k}}\otimes \upsilon _{j_{1}\cdots
j_{\ell }}
\end{equation}

Then, a calculation yields the reduced density matrix $\rho _{i_{m}}$\ for
qubit $i_{m}$, $m=1,\cdots ,k$,
\begin{equation}
\rho _{i_{m}}=tr_{(1,\cdots ,n)/i_{m}}\rho _{1\cdots n}=tr_{(i_{1},\cdots
,i_{k})/i_{m}}\sigma _{i_{1}\cdots i_{k}}=\sigma _{i_{m}}
\end{equation}%
and the reduced density matrix $\rho _{j_{m}}$ for qubit $j_{m}$, $%
m=1,\cdots ,\ell $,%
\begin{equation}
\rho _{j_{m}}=tr_{(1,\cdots ,n)/j_{m}}\rho _{1\cdots n}=tr_{(j_{1},\cdots
,j_{\ell })/j_{m}}\upsilon _{j_{1}\cdots j_{\ell }}=\upsilon _{j_{m}}
\end{equation}

Then,
\begin{eqnarray}
&&E_{AD}(|\psi \rangle _{1\cdots n}) \\
&=&\frac{1}{n}[4(\det \rho _{i_{1}}+\cdots +\det \rho _{i_{k}})+4(\det \rho
_{j_{1}}+\cdots +\det \rho _{j_{\ell }})] \\
&=&\frac{1}{n}[4(\det \sigma _{i_{1}}+\cdots +\det \sigma _{i_{k}})+4(\det
\upsilon _{j_{1}}+\cdots +\det \upsilon _{j_{\ell }})] \\
&=&\frac{kE_{AD}(|\phi \rangle _{i_{1}\cdots i_{k}})+\ell E_{AD}(|\varphi
\rangle _{j_{1}\cdots j_{\ell }})}{n}
\end{eqnarray}

Proposition 1 implies the following corollaries.

Corollary 1. If $|\psi \rangle _{1\cdots n}=|\phi \rangle _{i_{1}}\otimes
\cdots \otimes |\phi \rangle _{i_{k}}\otimes |\varphi \rangle _{rest}$,
where $|\phi \rangle _{i_{1}},\cdots ,|\phi \rangle _{i_{k}}$ are one-qubit
states, then $E_{AD}(|\psi \rangle _{1\cdots n})\leq \frac{n-k}{n}$.

Corollary 2. (i). If $E_{AD}(|\psi \rangle _{1\cdots n})=1$, then $|\psi
\rangle _{1\cdots n}$ is genuinely entangled or $|\psi \rangle _{1\cdots n}$
can be written as $|\psi \rangle _{1\cdots n}=|\phi \rangle _{i_{1}\cdots
i_{k}}\otimes \cdots \otimes |\phi \rangle _{j_{1}\cdots j_{\ell }}$, where $%
|\phi \rangle _{i_{1}\cdots i_{k}}$, $\cdots $, $|\phi \rangle _{j_{1}\cdots
j_{\ell }}$ are genuinely entangled and $E_{AD}(|\phi \rangle _{i_{1}\cdots
i_{k}})=\cdots =E_{AD}(|\phi \rangle _{j_{1}\cdots j_{\ell }})=1$.

(ii). Let $|\psi \rangle _{1\cdots n}=|\phi \rangle _{i_{1}\cdots
i_{k}}\otimes |\varphi \rangle _{j_{1}\cdots j_{\ell }}$, where\ $k+\ell =n$
and$\ |\phi \rangle _{i_{1}\cdots i_{k}}$ and $|\varphi \rangle
_{j_{1}\cdots j_{\ell }}$ are normalized. If $E_{AD}(|\phi \rangle
_{i_{1}\cdots i_{k}})=E_{AD}(|\varphi \rangle _{j_{1}\cdots j_{\ell }})=1$,
then $E_{AD}(|\psi \rangle _{1\cdots n})=1$ by the decomposition law.

For example, let $|$Bell$\rangle =$ $\frac{1}{\sqrt{2}}(|00\rangle
+|11\rangle )$. Then, $E_{AD}(|$Bell$\rangle )=1$. By the decomposition law,
$E_{AD}(|0\rangle \otimes |$Bell$\rangle )=2/3$ and $E_{AD}(|$Bell$\rangle
^{\otimes m})=1$.

\subsection{$E_{AD}$ is the average 1-tangle}

In \cite{Coffman}, for three qubits, 1-tangle $\tau _{i(jk)}$ is defined as $%
4\det \rho _{i}$. Therefore, for three qubits,
\begin{equation}
E_{AD}(|\psi \rangle )=\frac{1}{3}(\tau _{1(23)}+\tau _{2(13)}+\tau _{3(12)})
\end{equation}

That is, for three qubits, $E_{AD}(|\psi \rangle )$ is just the average
1-tangle, i.e. the average entanglement between one qubit and other two
qubits.

For $n$ qubits, 1-tangle $\tau _{i(1\cdots (i-1)(i+1)\cdots n)}$ can also be
defined as $4\det \rho _{i}$. Thus,

\begin{equation}
E_{AD}(|\psi \rangle )=\frac{1}{n}(\tau _{1(2\cdots n)}+\tau _{2(13\cdots
n)}+\cdots +\tau _{n(1\cdots (n-1))})  \label{1-tangle}
\end{equation}

Eq. (\ref{1-tangle}) means that $E_{AD}(|\psi \rangle )$ is just the average
1-tangle. That is, $E_{AD}(|\psi \rangle )$ is the average entanglement
between one qubit and the rest qubits.

\subsection{$E_{AD}$ is the average mixedness for each qubit}

It is known that the single-qubit state$\ \rho _{i}$ is the maximally mixed
state if it is proportional to the identity \cite{Scott, Gour}. In Appendix
B, we show that $\det \rho _{i}=1/4$ if and only if $\rho _{i}=(1/2)I_{2}$,
i.e. $\rho _{i}$ is proportional to $I_{2}$. So, when $\det \rho _{i}=1/4$,
by the definition $\rho _{i}$ is the maximally mixed state. Thus, $\det \rho
_{i}$ can be considered a measure of the mixedness of the single-qubit state
$\rho _{i}$ and $E_{AD}$ is the average mixedness for each qubit.

In \cite{Gour}, the absolutely maximally entangled (AME) state is defined as
the one whose reduced density matrices obtained by tracing out of any $k$
qubits, with $n/2\leq k\leq n-1$, are proportional to the identity. By the
definition, for the AME states, clearly $E_{AD}=1$. \footnote{%
It is known that the codeword states of the 5-qubit error correcting codes
are AME states \cite{Bennett, Laf, Gour}. For the codeword states, $E_{MW}=1$
\cite{Meyer}}

\subsection{$E_{AD}$ suggests not to use $n$-qubit W state or $n$-qubit
Dicke states with the fixed number of excitations (or \textquotedblleft
1\textquotedblright s) for the quantum system with the large number of qubits%
}

Let $|k,n\rangle $ stand for Dicke state of $n$ qubits, which is a uniform
superposition of all basis states with a fixed number of excitations (or
\textquotedblleft 1\textquotedblright s), $k$. It is known that $|1,n\rangle
$ is just W state of $n$ qubits. A calculation yields that $%
E_{AD}(|1,n\rangle )=4\frac{1}{n}(1-\frac{1}{n})$ and $E_{AD}(|k,n\rangle )=4%
\frac{k}{n}(1-\frac{k}{n})$.

\subsubsection{$E_{AD}(|k,n\rangle )$ decreases as $n$ increases for $n$%
-qubit W and Dicke states.}

When $k$ is fixed, $\lim_{n\rightarrow \infty }$ $E_{AD}(|k,n\rangle )=0$
and $E_{AD}(|k,n\rangle )$ decreases as $n$ increases. It means that $E_{AD}$
almost vanishes for large $n$ and the fixed $k$. For example, for $n$-qubit
W state, when $n=100$, $E_{AD}(|1,100\rangle )=$ $0.039\,6$. It suggests not
to use $n$-qubit W state or $n$-qubit Dicke states with the fixed number of
excitations (or \textquotedblleft 1\textquotedblright s) for the quantum
system with the large number of qubits whenever the average mixedness for
each qubit and 1-tangle are strongly required. So far, no one has proposed
this suggestion.

\subsubsection{When $n$ is fixed, $E_{AD}(|k,n\rangle )$ has the minimum at $%
n$-qubit W state and the maximum.}

When $n$ is fixed, $E_{AD}(|k,n\rangle )$ increases as $k$ does from 1 to $%
n/2$ for even $n$ (to $(n-1)/2$ for odd $n$) while $E_{AD}(|k,n\rangle )$
decreases as $k$ increases from $n/2$ (from $(n+1)/2$) to $n$ for even $n$
(for odd $n$). One can see when $n$ is fixed, $E_{AD}$ has the minimum of $4%
\frac{1}{n}(1-\frac{1}{n})$\ at the state $|1,n\rangle $ (i.e. $n$-qubit W
state), the maximum of $1$ at the state $|n/2,n\rangle $\ for even $n$, and
the maximum of $1-\frac{1}{n^{2}}$ at the states $|(n-1)/2,n\rangle $ and $%
|(n+1)/2,n\rangle $ for odd $n$.

\subsubsection{$E_{AD}$ also tends to zero as $n$ does to infinity for some
other states.}

For the following state $|\Psi \rangle $, $E_{AD}$ also tends to zero as $n$
does to infinity. Let the $n$-qubit state $|\Psi \rangle =\frac{1}{\sqrt{%
2^{n}}}\left( \sum_{i=0}^{2^{n}-2}|i\rangle -|2^{n}-1\rangle \right) $. $%
|\Psi \rangle $ can also be written as $\frac{1}{\sqrt{2^{n}}}(|0...0\rangle
+...+|1...10\rangle -|1...1\rangle )$. A calculation yields $E_{AD}(|\Psi
\rangle )=4\left[ \frac{1}{4}-\left( \frac{1}{2}-\frac{1}{2^{n-1}}\right)
^{2}\right] $. Clearly, $E_{AD}(|\Psi \rangle )$ decreases as $n$ increases
and $\lim_{n\rightarrow \infty }$ $E_{AD}(|\Psi \rangle )=0$. When $n=3$, $%
E_{AD}=3/4$. Thus, for $|\Psi \rangle $\ the mixedness for each qubit and
1-tangle almost vanish for the large number of qubits.

\subsection{Some conclusions for $E_{AD}$}

By means of the properties of the reduced density matrices $\rho _{i}$ and
from the above discussions, it is clear that the following Theorem 1 holds

Theorem 1. (i) $0\leq E_{AD}\leq 1$. (ii) $E_{AD}=1$ if and only if $\det
\rho _{i}=1/4$, i.e. $\rho _{i}$ is the maximally mixed state, $i=1,\cdots
,n $. (iii). $E_{AD}=0$ if and only if $\det \rho _{i}=0$, $\ i=1,\cdots ,n$%
, if and only if the state $|\psi \rangle $ is a fully separable state.
(iv). For biseparable states, i.e. not genuinely entangled or fully
separable states, $0<E_{AD}\leq 1$. Thus, that $E_{AD}=1$ means that the
state is genuinely entangled or biseparable. (v). $E_{AD}$ is LU invariant.

Clearly, (iii) implies (iv). We only prove (iii) below. $\det \rho _{i}=0$
means that qubit $i$ is not entangled with any other qubits in the system.

\section{Compare $E_{AD}$ to Meyer and Wallach's global entanglement measure}

\subsection{Meyer and Wallach's global entanglement measure}

Let $|\psi \rangle _{1\cdots n}=\sum_{i=0}^{2^{n}-1}c_{i}|i\rangle _{1\cdots
n}$ be any normalized pure state of $n$ qubits. We can write
\begin{equation}
|\psi \rangle _{1\cdots n}=|0\rangle _{k}|u^{(k)}\rangle +|1\rangle
_{k}|v^{(k)}\rangle ,  \label{mw-1}
\end{equation}%
where $|u^{(k)}\rangle $ and $|v^{(k)}\rangle $\ stand for the
non-normalized vectors $|u^{(k)}\rangle _{1\cdots (k-1)(k+1)\cdots n}$ and $%
|v^{(k)}\rangle _{1\cdots (k-1)(k+1)\cdots n}$, respectively, which are
called the projections of the state onto the $kth$ qubit subspaces \cite%
{Meyer, Roya}. We can also write $|u^{(k)}\rangle =_{k}\langle 0|\psi
\rangle _{1\cdots n}$ and $|v^{(k)}\rangle =_{k}\langle 1|\psi \rangle
_{1\cdots n}$.

In \cite{Meyer}, Meyer and Wallach proposed the following global
entanglement for pure states of $n$ qubits.

\begin{equation}
E_{MW}(|\psi \rangle )=\frac{4}{n}\sum_{k=1}^{n}D(|u^{(k)}\rangle
,|v^{(k)}\rangle ),
\end{equation}%
where $D(|u^{(k)}\rangle ,|v^{(k)}\rangle )$ is the norm-squared of the
wedge product of the two vectors $|u^{(k)}\rangle $ and $|v^{(k)}\rangle $
\begin{equation}
D(|u^{(k)}\rangle ,|v^{(k)}\rangle
)=\sum_{i<j}|u_{i}^{(k)}v_{j}^{(k)}-u_{j}^{(k)}v_{i}^{(k)}|^{2}.
\end{equation}

In \cite{Meyer}, they proved that $0\leq E_{MW}\leq 1$ and $E_{MW}=0$ if and
only if the state is fully separable.

$E_{AD}(|\psi \rangle )$ is the average determinant of reduced density
matrices for each qubit while $E_{MW}(|\psi \rangle )$ is the average
norm-squared of the wedge product of the two vectors $|u^{(k)}\rangle $ and $%
|v^{(k)}\rangle $ for each qubit. One can see that via the wedge product,\
it is not intuitive to propose the decomposition law, the mixedness or
1-tangle. We next show that $E_{MW}=E_{AD}$ \ algebraically\ for $n$ qubits
though $E_{AD}(|\psi \rangle )$ and $E_{MW}(|\psi \rangle )$ use different
concepts.

\subsection{$E_{AD}=E_{MW}$ for $n$ qubits}

Theorem 2. $D(|u^{(k)}\rangle ,|v^{(k)}\rangle )=\det \rho _{k}$, $%
k=1,2,\cdots ,n$. Thus, $E_{MW}=E_{AD}$.

We prove Theorem 2 for $n=2,3$ and any $n$ below.

\subsubsection{For two qubits}

We show that $D(|u^{(i)}\rangle ,|v^{(i)}\rangle )=\det \rho _{i}$, $i=1,2$,
and $E_{MW}=E_{AD}$ $=4|c_{0}c_{3}-c_{1}c_{2}|^{2}$\ in Appendix C. Then, we
conclude the following.

Proposition 2. For two qubits, $E_{MW}=E_{AD}$ and $E_{MW}$ and $E_{AD}$\
are just the square of the concurrence.

Werner states are defined as $\rho _{W}(p)=p|\psi _{B}\rangle \langle \psi
_{B}|+\frac{1-p}{4}I$, where $|\psi _{B}\rangle $ is Bell state, $I$ is the
identity, and $p\in \lbrack 0,1]$. The concurrence for $\rho _{W}$ is $%
C(\rho _{W})=\max \{0,\frac{3p-1}{2}\}$. Possibly, one can compute $E_{AD}$
for mixed states. It is interesting to compute $E_{AD}$ for $\rho _{W}$ and
compare $E_{AD}$ with the concurrence for $\rho _{W}$.

\subsubsection{For three qubits}

Let $|\psi \rangle _{123}=\sum_{i=0}^{7}c_{i}|i\rangle $ be any pure state
of three qubits. In Appendix D, we show $D(|u^{(1)}\rangle ,|v^{(1)}\rangle
)=\det \rho _{1}$. Similarly, we can show that $D(|u^{(k)}\rangle
,|v^{(k)}\rangle )=\det \rho _{k}$, $k=2,3$. Therefore, we can conclude $%
E_{MW}=E_{AD}$ for three qubits.

The average 2-tangles and the average square of the concurrences were
studied in \cite{Dur, li-qip}. We next compare $E_{AD}$\ with 3-tangle, the
average 2-tangles, and the average square of the concurrences. \ From
Appendix D and \cite{li-qip}, obtain
\begin{eqnarray}
E_{AD} &=&2\frac{\tau _{12}+\tau _{13}+\tau _{23}}{3}+\tau _{123}
\label{3-qubit} \\
&=&2\frac{C_{12}^{2}+C_{13}^{2}+C_{23}^{2}}{3}+\tau _{123}
\end{eqnarray}

Note that $\tau _{123}$ is 3-tangle. Then, we conclude the following
Proposition 3.

Proposition 3. $E_{AD}$ is the sum of 3-tangle and the double of the average
2-tangles. Thus, $E_{AD}$ is or greater than 3-tangle, the average
2-tangles, and the average square of the concurrences.

From \cite{li-qic} and Theorem 1, we can show that the following Proposition
4 holds.

Proposition 4. For three qubits, $E_{AD}=1$ (max) if and only if the state
is GHZ state under LU.

It is known that GHZ state is the maximally entangled state by several
measures. Proposition 4 implies that GHZ state is a unique maximally
entangled state by $E_{AD}$ under LU.

\subsubsection{For n qubits}

Let $|\psi \rangle _{12\cdots n}=\sum_{i=0}^{2^{n}-1}c_{i}|i\rangle $ be any
pure state of $n$ qubits. In Appendix E, we show that $\det \rho _{1}=D($ $%
|u^{(1)}\rangle $, $|v^{(1)}\rangle )$. Similarly, we can show $\det \rho
_{i}=D($ $|u^{(i)}\rangle $, $|v^{(i)}\rangle )$, $i=2,\cdots ,n$. Thus,
obtain $E_{MW}=E_{AD}$ for $n$ qubits.

\section{Comparing $E_{AD}$\ to von Neumann entropy and the linear entropy}

\subsection{Comparing $E_{AD}$\ to von Neumann entropy}

von Neumann entropy is defined as
\begin{equation}
S(\rho )=-\sum \eta _{i}\ln \eta _{i},  \label{von-1}
\end{equation}%
where $\eta _{i}\geq 0$ are the eigenvalues of $\rho $, and $\sum_{i}\eta
_{i}=1$.

By the second order Taylor expansion of $\ln (1\pm x)$, we can approximate $%
S(\rho _{i})$ as follows \cite{li-qip},

\begin{equation}
2S(\rho _{i})\approx 2\ln 2-1+4\det \rho _{i}  \label{app-1}
\end{equation}%
Let $E_{S}=$ $\frac{1}{n}\sum_{i=1}^{n}S(\rho _{i})$ be the average von
Neumann entropy for each qubit. Then,

\begin{equation}
2E_{S}\approx (2\ln 2-1)+E_{AD}
\end{equation}%
Thus, $E_{S}$ and $E_{AD}$ almost are linearly related.

\subsection{Comparing $E_{AD}$\ to the linear entropy}

\textit{\ }For any Hermitian 2 by 2 matrix $\rho $ with the trace of 1, it
satisfies

\begin{equation}
4\det \rho =2(1-\mathrm{Tr}(\rho ^{2}))  \label{linear-1}
\end{equation}

Thus, obtain
\begin{equation}
E_{AD}(|\psi \rangle )=\frac{1}{n}\sum_{i=1}^{n}2(1-\mathrm{Tr}(\rho
_{i}^{2})).  \label{linear-2}
\end{equation}

In \cite{Osborne}, the linear entropy $S_{2}(\rho _{i})$ for the
single-qubit state $\rho _{i}$ is defined as
\begin{equation}
S_{2}(\rho _{i})=2(1-\mathrm{Tr}(\rho _{i}^{2})).  \label{def-1}
\end{equation}

Then, obtain the following
\begin{equation}
E_{AD}(|\psi \rangle )=\frac{1}{n}\sum_{i=1}^{n}S_{2}(\rho _{i})
\end{equation}%
Therefore, $E_{AD}(|\psi \rangle )$ can also be called the average linear
entropy for each qubit.

In \cite{Brennen}, it was claimed that $E_{MW}=\frac{1}{n}\sum_{i=1}^{n}2(1-%
\mathrm{Tr}(\rho _{i}^{2}))$, i.e. $E_{MW}$ is the linear entropy, which is
extended to the general case \cite{Scott}. This claim was derived via the
condition $\langle \tilde{x}^{k}|\tilde{y}^{k}\rangle =0$ \cite{Brennen}. We
deduce that $E_{AD}=E_{MW}$. Thus, our proof for that $E_{AD}$ and $E_{MW}$
both are the linear entropy is different from the one \cite{Brennen}.

Remark 1. One can check that for the single-qubit state $\rho _{i}$, $4\det
\rho _{i}=(4/3)(1-\mathrm{Tr}(\rho _{i}^{3}))$. So, the linear entropy $%
S_{2}(\rho _{i})$ can also be defined as $S_{2}(\rho _{i})=(4/3)(1-\mathrm{Tr%
}(\rho _{i}^{3}))$.

\section{Discussion}

Note that $E_{AD}=1$ for some biseparable states of $n$($\geq 4$) qubits,
for example for $|\Phi ^{+}\rangle $. Thus, that $E_{AD}=1$ (max) does not
imply the state is genuinely entangled. It means that $E_{AD}=1$ can not
distinguish biseparable states and genuinely entangled states. To overcome
the weakness, we need to compute all the reduced density matrices obtained
by tracing out of any $m$ qubits, with $n/2\leq m\leq n-1$. Let $\kappa
=\left(
\begin{array}{c}
n \\
\ell%
\end{array}%
\right) $ and
\[
E_{AD}^{(\ell )}=\frac{\mu }{\kappa }\sum_{i_{1}\cdots i_{\ell }}\det \rho
_{i_{1}\cdots i_{\ell }},1\leq \ell \leq n/2,
\]%
where the constant $\mu $ makes the normalization for $E_{AD}^{(\ell )}$.
Then, let $E_{AD}^{(g)}$ be the average of $E_{AD}^{(\ell )}$, $1\leq \ell
\leq n/2$. By the properties of reduced density matrices it is easy to
obtain (i). $0\leq E_{AD}^{(g)}\leq 1$. (ii) $E_{AD}^{(g)}=1$ if and only if
the state is AME state. (iii). $E_{AD}^{(g)}=0$ if and only if the state $%
|\psi \rangle $ is a fully separable state. (iv). For biseparable states, $%
0<E_{AD}<1$. (v). $E_{AD}^{(g)}$ is LU invariant.

Remark 2. For $n=4$ and $n\geq 7$, AME states don't exist \cite{Gour, Huber}%
, while for $n=3,5,6$, the AME states exist \cite{Bennett, Laf, Rains, Huber}%
. For example, three-qubit GHZ state is the AME state.

\section{Summary}

In this paper, we propose the global entanglement measure $E_{AD}$. $E_{AD}$
measures the \ average mixedness of quantum states for each qubit and the
average entanglement between one qubit and the rest qubits. We present the
decomposition law for $E_{AD}$. So far no one has proposed it.

We compare $E_{AD}$\ to Meyer-Wallach's global entanglement measure,\ von
Neumann entropy, and the linear entropy. We show that $E_{AD}$ is just
Meyer-Wallach's global entanglement measure $E_{MW}$ by straightforwardly
calculating $E_{AD}$ and $E_{MW}$. Clearly, it is not intuitive to propose
the decomposition law, the mixedness or 1-tangle via the wedge products.

\section{Appendix A. Computing $E_{AD}$ for some states}

We can compute $E_{AD}$ for the GHZ-like states of $n$ qubits $\alpha
|0\cdots 0\rangle +\beta |1\cdots 1\rangle $, where $\alpha ,\beta >0$ and $%
\alpha ^{2}+\beta ^{2}=1$. It is easy to see that $\rho _{i}=diag(\alpha
^{2},\beta ^{2})$, $i=1,\cdots ,n$. Then, $E_{AD}=4\alpha ^{2}\beta ^{2}$.
Specially, when $\alpha =\beta =1/\sqrt{2}$, i.e. the $n$-qubit GHZ state, $%
E_{AD}=1$.

A calculation yields $E_{AD}=1$ for the following states $|S_{n}\rangle $
and $|\Phi ^{\pm }\rangle $ of $n$-qubits and the two graph states of three
qubits. We define the following symmetric state $|S_{n}\rangle $\ of even n
qubits. Let $i_{1}i_{2}\cdots i_{n}$ be an n-bit binary number, $%
i_{j}^{\prime }$ be the complement of $i_{j}$, and $\ell $ be the number of
\textquotedblleft 1\textquotedblright s in $i_{1}i_{2}\cdots i_{n}$. Let

\begin{eqnarray}
|S_{n}\rangle &=&c_{0\cdots 0}(|0\cdots 0\rangle +|1\cdots 1\rangle )
\nonumber \\
&&+\sum_{i_{1},\cdots i_{n}=0,1,\ell =n/2}c_{i_{1}i_{2}\cdots
i_{n}}(|i_{1}i_{2}\cdots i_{n}\rangle +|i_{1}^{\prime }i_{2}^{\prime }\cdots
i_{n}^{\prime }\rangle ).
\end{eqnarray}%
Specially for four qubits, $|S_{n}\rangle $ is reduced to $G_{abcd}$ \cite%
{Verstraete}.

For even $n$ qubits, let
\begin{eqnarray}
|\Phi ^{\pm }\rangle &=&\frac{1}{2}(|0\cdots 0\rangle _{1\cdots n}+|0\cdots
0\rangle _{1\cdots (n/2)}|1\cdots 1\rangle _{(n/2+1)\cdots n}  \nonumber \\
&&+|1\cdots 1\rangle _{1\cdots (n/2)}|0\cdots 0\rangle _{(n/2+1)\cdots n}\pm
|1\cdots 1\rangle _{1\cdots n}),
\end{eqnarray}

Note that $|\Phi ^{-}\rangle $ is called the cluster state while $|\Phi
^{+}\rangle $ is biseparable.

The following are the two graph states of three qubits.

\begin{eqnarray}
&&\frac{1}{2\sqrt{2}}(|000\rangle +|001\rangle +|010\rangle -|011\rangle
+|100\rangle +|101\rangle -|110\rangle +|111\rangle ),  \label{graph-1} \\
&&\frac{1}{2\sqrt{2}}(|000\rangle +|001\rangle +|010\rangle -|011\rangle
+|100\rangle -|101\rangle -|110\rangle -|111\rangle ).  \label{graph-2}
\end{eqnarray}

\section{ Appendix B. Mixedness}

Result 1. $\det \rho _{i}=1/4$ if and only if $\rho _{i}=(1/2)I_{2}$, i.e. $%
\rho _{i}$ is proportional to $I_{2}$.

Proof. Let the reduced density matrix $\rho _{i}=\left(
\begin{array}{cc}
a & b \\
c & d%
\end{array}%
\right) $. Then, $\rho _{i}$ is Hermitian and has the trace of 1. Thus, $a$
and $d$ are real, $c=b^{\ast }$, where $b^{\ast }$ is the complex conjugate
of $b$, and $a+d=1$.

Assume that $\det \rho _{i}=1/4$. Then, $ad-bc=ad-|b|^{2}=1/4$. From $%
ad-|b|^{2}=1/4$, one can know that $a$ and $d$ both are positive or
negative. Then, from $a+d=1$, it is easy to see that $a$ and $d$ both are
positive. It is also known that $ad\leq \left( \frac{a+d}{2}\right) ^{2}=%
\frac{1}{4}$. Then, from that $ad-|b|^{2}=1/4$, obtain $b=0$ and $ad=1/4$.
From that $ad=1/4$ and $a+d=1$, obtain $a=d=1/2$. Thus, $\rho
_{i}=(1/2)I_{2} $.

Conversely, it is trivial to see it holds.{}

\section{Appendix C. For two qubits}

Any pure state of two qubits can be written as $|\psi \rangle
_{12}=(c_{0}|00\rangle +c_{1}|01\rangle +c_{2}|10\rangle +c_{3}|11\rangle
)_{12}$. We can rewrite
\begin{eqnarray}
|\psi \rangle _{12} &=&|0\rangle _{1}(c_{0}|0\rangle +c_{1}|1\rangle
)_{2}+|1\rangle _{1}(c_{2}|0\rangle +c_{3}|1\rangle )_{2} \\
&=&|0\rangle _{2}(c_{0}|0\rangle +c_{2}|1\rangle )_{1}+|1\rangle
_{2}(c_{1}|0\rangle +c_{3}|1\rangle )_{1}
\end{eqnarray}

A calculation yields
\begin{equation}
D(|u^{(1)}\rangle ,|v^{(1)}\rangle )=D(|u^{(2)}\rangle ,|v^{(2)}\rangle
)=|c_{0}c_{3}-c_{1}c_{2}|^{2}.
\end{equation}

We next calculate $\det \rho _{i}$, $i=1,2$. It is known that $\rho
_{1}=C_{2}C_{2}^{H}$, where
\begin{equation}
C_{2}=\left(
\begin{array}{cc}
c_{0} & c_{1} \\
c_{2} & c_{3}%
\end{array}%
\right)
\end{equation}%
and $C_{2}^{H}$ is the Hermitian transpose of $C_{2}$. A calculation yields
\begin{eqnarray}
\det \rho _{1} &=&c_{0}c_{3}c_{0}^{\ast }c_{3}^{\ast }-c_{0}c_{3}\allowbreak
c_{1}^{\ast }c_{2}^{\ast }-c_{1}c_{2}c_{0}^{\ast }c_{3}^{\ast
}+c_{1}c_{2}c_{1}^{\ast }c_{2}^{\ast }  \label{2-q-1} \\
&=&|c_{0}c_{3}-c_{1}c_{2}|^{2}.
\end{eqnarray}

Note that $c_{i}^{\ast }$ is the complex conjugate of $c_{i}$. Similarly, $%
\det \rho _{2}=|c_{0}c_{3}-c_{1}c_{2}|^{2}$. Thus, $\det \rho
_{k}=D(|u^{(k)}\rangle ,|v^{(k)}\rangle )$, $k=1,2$, and then $E_{MW}=E_{AD}$
$=4|c_{0}c_{3}-c_{1}c_{2}|^{2}$.

\section{Appendix D. For three qubits}

For three qubits, let $|\psi \rangle _{123}=\sum_{i=0}^{7}c_{i}|i\rangle $.

\subsection{Calculate $D(|u^{(1)}\rangle ,|v^{(1)}\rangle )$}

By the definition of the vectors $|u^{(k)}\rangle $ and $|v^{(k)}\rangle $
\cite{Meyer}, obtain
\begin{eqnarray*}
|u^{(1)}\rangle &=&c_{0}|00\rangle +c_{1}|01\rangle +c_{2}|10\rangle
+c_{3}|11\rangle , \\
|v^{(1)}\rangle &=&c_{4}|00\rangle +c_{5}|01\rangle +c_{6}|10\rangle
+c_{7}|11\rangle .
\end{eqnarray*}

One can see that
\begin{equation}
D(|u^{(1)}\rangle ,|v^{(1)}\rangle
)=\sum_{i<j}|u_{i}^{(1)}v_{j}^{(1)}-u_{j}^{(1)}v_{i}^{(1)}|^{2}=%
\sum_{i<j}|c_{i}c_{4+j}-c_{j}c_{4+i}|^{2}.  \label{3-q-1}
\end{equation}

\subsection{Calculate $\det \protect\rho _{1}$}

It is known that $\rho _{1}=tr_{23}\rho _{123}=C_{3}C_{3}^{H}$, where
\[
C_{3}=\left(
\begin{array}{cccc}
c_{0} & c_{1} & c_{2} & c_{3} \\
c_{4} & c_{5} & c_{6} & c_{7}%
\end{array}%
\right) .
\]%
A calculation yields that
\[
\rho _{1}=\left(
\begin{array}{cc}
\Delta _{1} & \Delta _{2} \\
\Delta _{3} & \Delta _{4}%
\end{array}%
\right) ,
\]%
where

\begin{eqnarray}
\Delta _{1} &=&\sum_{i=0}^{3}c_{i}c_{i}^{\ast },\Delta
_{2}=\sum_{i=0}^{3}c_{i}c_{4+i}^{\ast }, \\
\Delta _{3} &=&\sum_{i=0}^{3}c_{i}^{\ast }c_{4+i},\Delta
_{4}=\sum_{i=0}^{3}c_{4+i}^{\ast }c_{4+i}.
\end{eqnarray}

Clearly, $\det \rho _{1}=\Delta _{1}\Delta _{4}-\Delta _{2}\Delta _{3}$ and
\begin{equation}
\Delta _{1}\Delta _{4}-\Delta _{2}\Delta _{3}=\sum_{i\neq
j,i,j=0,1,2,3}\Theta _{ij},
\end{equation}%
where
\begin{equation}
\Theta _{ij}=c_{i}c_{i}^{\ast }c_{4+j}^{\ast }c_{4+j}-c_{i}c_{4+i}^{\ast
}c_{j}^{\ast }c_{4+j}.
\end{equation}

For example, $\Theta _{01}=c_{0}c_{0}^{\ast }c_{5}^{\ast
}c_{5}-c_{0}c_{4}^{\ast }c_{1}^{\ast }c_{5}$ and $\Theta
_{10}=c_{1}c_{1}^{\ast }c_{4}^{\ast }c_{4}-c_{1}c_{5}^{\ast }c_{0}^{\ast
}c_{4}$. Then, $\Theta _{01}+\Theta _{10}=|c_{0}c_{5}-c_{1}c_{4}|^{2}$.

Generally, when $i\neq j$,
\begin{eqnarray}
&&\Theta _{ij}+\Theta _{ji}  \nonumber \\
&=&(c_{i}c_{i}^{\ast }c_{4+j}^{\ast }c_{4+j}-c_{i}c_{4+i}^{\ast }c_{j}^{\ast
}c_{4+j})+(c_{j}c_{j}^{\ast }c_{4+i}^{\ast }c_{4+i}-c_{j}c_{4+j}^{\ast
}c_{i}^{\ast }c_{4+i})  \nonumber \\
&=&(c_{i}c_{4+j}-c_{j}c_{4+i})(c_{i}^{\ast }c_{4+j}^{\ast }-c_{j}^{\ast
}c_{4+i}^{\ast })  \nonumber \\
&=&|c_{i}c_{4+j}-c_{j}c_{4+i}|^{2}.
\end{eqnarray}

Thus,

\begin{equation}
\det \rho _{1}=\sum_{i\neq j,i,j=0,1,2,3}\Theta
_{ij}=\sum_{i<j}|c_{i}c_{4+j}-c_{j}c_{4+i}|^{2}.  \label{3-q-2}
\end{equation}%
Therefore, from Eqs. (\ref{3-q-1}, \ref{3-q-2}), obtain $D(|u^{(1)}\rangle
,|v^{(1)}\rangle )=\det \rho _{1}$.

\section{Appendix E. For n qubits}

Let $|\psi \rangle _{12\cdots n}=\sum_{i=0}^{2^{n}-1}c_{i}|i\rangle $ be any
pure state of $n$ qubits. We show that $\det (\rho _{1})=D($ $%
|u^{(1)}\rangle $, $|v^{(1)}\rangle )$ below.

\subsection{Calculating $\det \protect\rho _{1}$}

By the definition, $\rho _{1}=tr_{23...n}\rho _{12...n}$. One can see that $%
\rho _{1}=C_{n}C_{n}^{H}$, where
\begin{equation}
C_{n}=\left(
\begin{array}{ccccc}
c_{0} & c_{1} & \cdots & c_{2^{n-1}-2} & c_{2^{n-1}-1} \\
c_{2^{n-1}} & c_{2^{n-1}+1} & \cdots & c_{2^{n}-2} & c_{2^{n}-1}%
\end{array}%
\right) .
\end{equation}

A calculation yields that
\[
\rho _{1}=\left(
\begin{array}{cc}
\Delta _{1} & \Delta _{2} \\
\Delta _{3} & \Delta _{4}%
\end{array}%
\right) ,
\]%
where

\begin{eqnarray}
\Delta _{1} &=&\sum_{i=0}^{2^{n-1}-1}c_{i}c_{i}^{\ast },\Delta
_{2}=\sum_{i=0}^{2^{n-1}-1}c_{i}c_{2^{n-1}+i}^{\ast }, \\
\Delta _{3} &=&\sum_{i=0}^{2^{n-1}-1}c_{i}^{\ast }c_{2^{n-1}+i},\Delta
_{4}=\sum_{i=0}^{2^{n-1}-1}c_{2^{n-1}+i}^{\ast }c_{2^{n-1}+i}.
\end{eqnarray}%
$\allowbreak $ $\allowbreak $

Then,
\begin{equation}
\det \rho _{1}=\Delta _{1}\Delta _{4}-\Delta _{2}\Delta _{3}=\sum_{i\neq
j,i,j\in \{0,1,\cdots ,(2^{n-1}-1)\}}\digamma _{ij},
\end{equation}%
where
\begin{equation}
\digamma _{ij}=c_{i}c_{i}^{\ast }c_{2^{n-1}+j}c_{2^{n-1}+j}^{\ast
}-c_{i}c_{j}^{\ast }c_{2^{n-1}+i}^{\ast }c_{2^{n-1}+j}.
\end{equation}

Generally, when $i\neq j$,%
\begin{eqnarray}
\digamma _{ij}+\digamma _{ji} &=&c_{i}c_{i}^{\ast
}c_{2^{n-1}+j}c_{2^{n-1}+j}^{\ast }-c_{i}c_{j}^{\ast }c_{2^{n-1}+i}^{\ast
}c_{2^{n-1}+j}  \nonumber \\
&&+c_{j}c_{j}^{\ast }c_{2^{n-1}+i}c_{2^{n-1}+i}^{\ast }-c_{j}c_{i}^{\ast
}c_{2^{n-1}+j}^{\ast }c_{2^{n-1}+i}  \nonumber \\
&=&(c_{i}c_{2^{n-1}+j}-c_{j}c_{2^{n-1}+i})(c_{i}^{\ast }c_{2^{n-1}+j}^{\ast
}-c_{j}^{\ast }c_{2^{n-1}+i}^{\ast })  \nonumber \\
&=&|c_{i}c_{2^{n-1}+j}-c_{j}c_{2^{n-1}+i}|^{2}
\end{eqnarray}

Then,
\begin{equation}
\sum_{i\neq j,i,j\in \{0,1,\cdots ,(2^{n-1}-1)\}}\digamma
_{ij}=\sum_{i<j}|c_{i}c_{2^{n-1}+j}-c_{j}c_{2^{n-1}+i}|^{2}.
\end{equation}%
Thus, we obtain
\begin{equation}
\det \rho _{1}=\sum_{i<j}|c_{i}c_{2^{n-1}+j}-c_{j}c_{2^{n-1}+i}|^{2}
\label{eq-1}
\end{equation}

\subsection{Calculating $D($ $|u^{(1)}\rangle $, $|v^{(1)}\rangle )$}

We calculate $\sum_{i<j}|u_{i}^{(1)}v_{j}^{(1)}-u_{j}^{(1)}v_{i}^{(1)}|^{2}$
below. We can write

\begin{equation}
|\psi \rangle _{12...n}=|0\rangle _{1}|u^{(1)}\rangle _{2\cdots n}+|1\rangle
_{1}|v^{(1)}\rangle _{2\cdots n},
\end{equation}

where
\begin{eqnarray}
|u^{(1)}\rangle _{2\cdots n} &=&(c_{0}|0\rangle +c_{1}|1\rangle
+...+c_{2^{n-1}-1}|2^{n-1}-1\rangle )_{2\cdots n}, \\
|v^{(1)}\rangle _{2\cdots n} &=&(c_{2^{n-1}}|0\rangle
+c_{2^{n-1}+1}|1\rangle +...+c_{2^{n}-1}|2^{n-1}-1\rangle )_{2...n},
\end{eqnarray}

Then, we list the following coefficients of the vectors $|u^{(1)}\rangle
_{2\cdots n}$ and $|v^{(1)}\rangle _{2\cdots n}$
\begin{equation}
(u_{0}^{(1)},u_{1}^{(1)},\cdots ,u_{2^{n-1}-1}^{(1)})=(c_{0},c_{1},\cdots
,c_{2^{n-1}-1})  \label{coe-5}
\end{equation}%
and
\begin{equation}
(v_{0}^{(1)},v_{1}^{(1)},\cdots
,v_{2^{n}-1}^{(1)})=(c_{2^{n-1}},c_{2^{n-1}+1},\cdots ,c_{2^{n}-1})
\label{coe-6}
\end{equation}

From Eqs. (\ref{coe-5}, \ref{coe-6}), for $i$ and $j$,\ $u_{i}^{(1)}=c_{i}$,
$v_{j}^{(1)}=c_{2^{n-1}+j}$, $u_{j}^{(1)}=c_{j}$, and $%
v_{i}^{(1)}=c_{2^{n-1}+i}$. Then,%
\begin{equation}
|u_{i}^{(1)}v_{j}^{(1)}-u_{j}^{(1)}v_{i}^{(1)}|^{2}=|c_{i}c_{2^{n-1}+j}-c_{j}c_{2^{n-1}+i}|^{2}.
\end{equation}

Thus,
\begin{equation}
D(|u^{(1)}\rangle ,|v^{(1)}\rangle
)=\sum_{i<j}|c_{i}c_{2^{n-1}+j}-c_{j}c_{2^{n-1}+i}|^{2}  \label{eq-2}
\end{equation}

Then, from Eqs. (\ref{eq-1}, \ref{eq-2}), we can obtain
\begin{equation}
\det (\rho _{1})=D(|u^{(1)}\rangle ,|v^{(1)}\rangle ).
\end{equation}

\section{The conflict of interest statement}

No conflict of interest.

\section{The data availability statement}

It includes all data in the main text. \

Acknowledgement

Thank Julia Huang (of Stanford University) for changing English.

\end{document}